\documentclass[fleqn,twoside,twocolumn,nofootinbib,showkeys]{revtex4} 
\usepackage[nocpr]{ujp} 

\begin{document}
\title[Momentum Diffusion of Atoms and Nanoparticles]
{MOMENTUM DIFFUSION\\ OF ATOMS AND NANOPARTICLES\\ IN AN OPTICAL
TRAP FORMED BY
SEQUENCES\\ OF COUNTER-PROPAGATING LIGHT PULSES}%
\author{V.I.~Romanenko}
\affiliation{Institute of Physics, Nat. Acad. of Sci. of Ukraine}
\address{46, Prosp. Nauky, Kyiv 03028, Ukraine}
\email{vr@iop.kiev.ua}
\author{A.V.~Romanenko}%
\affiliation{Taras Shevchenko National University of Kyiv}%
\address{4, Prosp. Academician Glushkov, Kyiv 03022, Ukraine}%
\email{alexrm@univ.kiev.ua}
\author{Ye.G.~Udovitskaya}
\affiliation{Institute of Physics, Nat. Acad. of Sci. of Ukraine}
\address{46, Prosp. Nauky, Kyiv 03028, Ukraine}
\email{vr@iop.kiev.ua}
\author{L.P.~Yatsenko}%
\affiliation{Institute of Physics, Nat. Acad. of Sci. of Ukraine}%
\address{46, Prosp. Nauky, Kyiv 03028, Ukraine}%
\email{vr@iop.kiev.ua}

\udk{535.214} \pacs{42.50.Wk} \razd{\seciv}

\autorcol{V.I.\hspace*{0.7mm}Romanenko,
A.V.\hspace*{0.7mm}Romanenko, Ye.G.\hspace*{0.7mm}Udovitskaya et
al.}

\setcounter{page}{438}%


\begin{abstract}
The motion of atoms and nanoparticles in a trap formed by sequences
of counter-propagating light pulses has been analyzed. The atomic
state is described by a wave function constructed with the use of the
Monte Carlo method, whereas the atomic motion is considered in the
framework of classical mechanics. The effects of the momentum
diffusion associated with the spontaneous radiation emission by
excited atoms and the pulsed character of the atom-to-field
interaction on the motion of a trapped atom or nanoparticle are
estimated. The motion of a trapped atom is shown to be slowed down
for properly chosen parameters of the atom-to-field interaction, so that
the atom oscillates around the antinodes of a non-stationary standing
wave formed by counter-propagating light pulses at the point where
they \textquotedblleft collide\textquotedblright.
\end{abstract}
\keywords{light pressure force, counter-propagating pulses,  trap,
nanoparticles, Monte Carlo wave function approach.}

\maketitle

\section{Introduction}

\label{introduction}Mechanical action of light on atoms
\cite{Min86,Kaz91,Pav93,Met99,Chu98,Coh98,Phi98,Bal00} is a cornerstone of
modern atomic optics. In most cases, the required strengths of a light pressure
are attained with the use of continuous laser radiation, which may be an
additional factor that decreases the accuracy of physical experiments owing to
light shifts induced by laser radiation. The control of the atomic motion with the
help of light pulses can be a promising alternative, which would enable the
interaction between the atom and the field to be so organized that the atom
would be subjected to the action of laser radiation only within a short time
intervals \cite{Fre95,Goe97,Bal05,Rom11,Rom12}.

The optical trap for atoms proposed in work \cite{Fre95} is based of
the interaction between atoms and sequences of counter-propagating
$\pi$-pulses, which was studied in works \cite{Kaz74,Voi91,Neg08}\
in detail. Figure~1 illustrates the mechanism of trap action. Let
light pulses propagate along the $z$-axis. An atom at point $A$ has
just undergone the action of pulse~$R$ propagating from left to
right, and soon it will be subjected to the action of pulse~$L$
propagating from right to left. If this atom was in the ground state
before the action of pulse~$R$, the interaction with the latter
transforms it into the excited state with the momentum $\hbar k$
directed along the $z$-axis. After being subjected to the action of
pulse~$L$, the atom emits a photon, and its momentum changes by
another $\hbar k$ in the same direction. As a result of the
interaction between the atom at point $A$ and a sequence of pulses
that repeat with the period $T$, the atom is subjected to the action
of the average force $2\hbar k/T$. A similar reasoning for an atom
at point~$B$~-- in this case, the atom is first subjected to the
action of pulse~$L$ and, afterward, pulse~$R$~-- brings us to a
conclusion that, owing to the interaction with a pair of
counter-propagating pulses, its momentum changes by $-2\hbar k$, so
that the average force acting on it equals $-2\hbar k/T$, i.e.
directed toward point~$C$. From the symmetry of the interaction
between the atom and the field at point~$C$, it follows that the
force of light pressure on the atom equals zero. Hence,
counter-propagating light $\pi$-pulses can form a trap for an atom.
As was marked in works \cite{Fre95,Goe97}, pulses with areas
different from $\pi$ can also be used for \mbox{this purpose.}

The basic idea of the trap~-- invoking such a light pressure force
that would act on the atom toward the point, where the
counter-propagating pulses \textquotedblleft
collide\textquotedblright~-- was used in work \cite{Goe97} in order to
focus a beam of atoms. The trap proposed in work \cite{Fre95} was
theoretically analyzed in works \cite{Fre95,Rom11}. The authors of
work \cite{Fre95} considered a variation of atom's momentum and
its scattering in the region of spatial overlapping between
counter-propagating light pulses, when the atom interacts only once
with them. The results obtained in the cited work are valid if two
conditions are satisfied. First, the pulse repetition period should
considerably exceed the lifetime of an atom in the excited state.
Second, the atom must be in a state described by a wide wave packet,
much wider than the wavelength of laser radiation. The authors of
work \cite{Fre95} proposed the laser cooling in order to compensate the
scattering-induced \textquotedblleft heating\textquotedblright\ of
atoms. In work \cite{Rom11}, the force of light pressure averaged
over an ensemble of slow atoms in such a trap was calculated for
light pulses of an arbitrary area. Hence, the scattering of atoms in
the region of spatial pulse overlapping, where the laser radiation
field is close to a standing wave, was omitted from consideration.

The frequency detuning of monochromatic counter-propagating waves from the
resonance with the frequency of an atomic transition is known \cite{Let76,Let77}
to bring about the so-called Doppler cooling of the atomic ensemble, if the field
frequency is lower than the transition one. The detuning magnitude should be
close to the reciprocal lifetime of an atom in the excited state. In this case,
the frequency of a transition in the atom, owing to the Doppler effect, is
closer to the resonance with the counter-propagating wave, irrespective of the
direction of atom's motion, and the force of light pressure that slows down
the atom exceeds the pressure force from the other wave that accelerates it.
As a result, the atomic ensemble is cooled down. In work \cite{Mol91}, it was
demonstrated that atoms can also be cooled down in the field of
counter-propagating pulses. Estimations of the light pressure force and the
coefficient of momentum diffusion were made in work \cite{Mol91} for
low-intensity fields, when they are equal to the sum of corresponding
contributions made by counter-propagating waves. At the same time, for the
analysis of the light trap in the field of counter-propagating pulses, high
intensities of fields are optimal, when the areas of light pulses are close to
$\pi$ and the force of light pressure cannot be considered as a sum of forces
associated with each of the running sequences of counter-propagating pulses.

\begin{figure}
\vskip1mm
\includegraphics[width=\column]{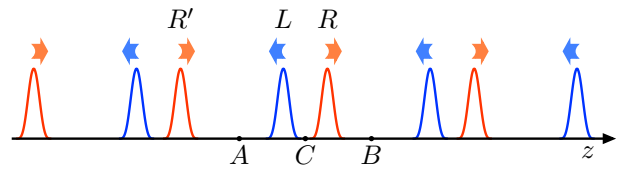}
\vskip-3mm\caption{Interaction between an atom and sequences of
counter-propagating pulses results in emerging a force that holds
the atom near point C, where the pulses \textquotedblleft
collide\textquotedblright  }\label{fig-scheme}
\end{figure}

In this work, we analyze the motion of an atom in a trap formed by light
$\pi$-pulses. By examining the resonance interaction between the atom and the
field, we shall estimate the influence of the momentum diffusion on atom's
motion in the trap. If the carrier frequency of light pulses is detuned from
that of an atomic transition, a decelerating force emerges under certain
conditions, and its influence can prevail over the influence of the momentum
diffusion. As a result, the amplitude of atom's oscillations in the trap
becomes smaller than the wavelength. At the same time, owing to the momentum
diffusion, the equilibrium position of the atom, around which it oscillates
(at antinodes of the field), occasionally changes by half a wavelength. We did not
manage to obtain analytical expressions for the description of this motion.
Therefore, our research is based on the numerical simulation. To describe the
evolution of the atomic state, we use a wave function constructed with the help of
the Monte Carlo method \cite{Mol93}. Atom's motion is described in the
framework of classical mechanics, which corresponds to a narrow, in comparison
with the wavelength, atomic wave packet.

In Section~\ref{sec:pulses}, the light fields that act on the atom
are considered. The model shape of pulses, which is introduced
there, is close to Gaussian-like, but is confined in time. In
Section~\ref{sec:Hamiltonian}, we describe the Hamiltonian, which is
used in Section~\ref{sec:WF} to construct the wave function with the
use of the Monte Carlo method. Expressions for the calculation of
the force and Newton's equations needed for the description of
atom's motion are presented in Section~\ref{sec:Motion}. The
procedure of numerical simulation of the state of atom and its
motion is described in Section~\ref{sec:num}. Section~\ref{sec:free}
illustrates the application of the wave function constructed with
the help of the Monte Carlo method to the description of a free,
i.e. in the absence of the field, evolution of the atomic state.
Here, the averaged elements of the density matrix for an ensemble of
atoms are compared with the known analytical expressions. The
results obtained at simulating the motion of atoms and nanoparticles
in a trap formed by sequences of counter-propagating light pulses
with the carrier frequency resonant with the frequency of a
transition in the atom are reported in Section~\ref{sec:pi}. In
Section~\ref{sec:det}, we substantiate a possibility to keep atoms
in the trap and, simultaneously, to cool them down by the same
field. At last, in Section~\ref{sec:conclu}, the conclusions of the
work are formulated in brief.\vspace*{-2mm}

\begin{figure}
\vskip1mm
\includegraphics[width=7.5cm]{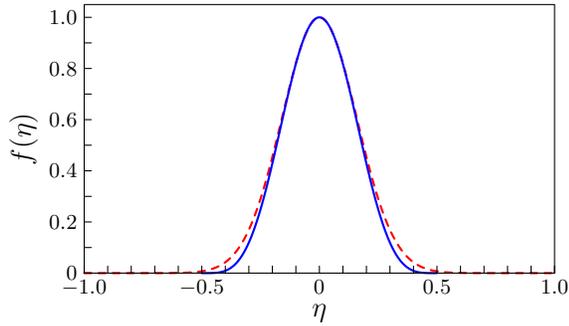}
\vskip-3mm\caption{Comparison of the function $f(\eta)$ (see
Eq.~(\ref{eq:fcos})) describing the envelope of light pulses (solid
curve) and a Gaussian function that is the closest to it (dashed
curve)  }\label{fig:pulse}\vspace*{-2mm}
\end{figure}

\section{Light Pulses}

\label{sec:pulses}Consider a two-level atom with the ground state
$|1\rangle$, excited state $|2\rangle$, and the frequency of the
transition between them $\omega_{0}$. The atom interacts with the
field created by two counter-propagating sequences of
pulses,\vspace*{-1mm}
\begin{equation}
\bm{\mathrm{E}}(t)\!=\!E_{1}(t)\bm{\mathrm{e}}\cos[\omega{}t-kz+\varphi_{1}%
]\!+\!E_{2}(t)\bm{\mathrm{e}}\cos[\omega{}t\!+\!kz\!+\!\varphi_{2}].\label{eq:E}%
\end{equation}\vspace*{-5mm}

\noindent Here, $E_{1,2}(t)$ are the pulse envelopes, $k=\omega/c$,
$\mathbf{e}$ is the unit vector of polarization for the electric
fields of pulses, $\varphi_{1}$ and $\varphi_{2}$ are the field
phases at $t=0$ and $z=0$. To simplify the notations, the arguments
for the field amplitudes, the density matrix elements, and the
probability amplitudes will be omitted in most cases. We consider
the interaction of an atom with a field created by two sequences of
counter-propagating pulses with the repetition period $T.$ At the
atom location point, one of the sequences repeats the other one with
a certain time delay. The amplitudes of both pulses are described by
the expression\vspace*{-1mm}
\begin{equation}
E_{1{,}2}=E_{0}f(\eta_{1,2}){,}
\end{equation}
where the function $f(\eta)$ with the maximum value $f(0)=1$ describes the
shape of a pulse envelope,
\begin{equation}
\eta_{1{,}2}=(t\mp{}z/c)/\tau{,}
\end{equation}
$z$ is the coordinate of the atom, and $\tau$ the pulse
\mbox{duration.}

When the interaction between an atom and a field is simulated, the
Gaussian-like pulses are usually applied \cite{Ber98,Vit01}. It is known
\cite{Rom06,Neg08} that the function $\cos^{n}(\pi t/\tau)$ for large even
$n$ tends to $\exp(-t^{2}/\tau_{\rm G}^{2})$ with $\tau_{\rm G}=\tau\sqrt{2}%
/(\pi\sqrt{n})$ within the interval $\left\vert t\right\vert <\tau/2$.
Therefore, let us take $f(\eta)$ in the form
\begin{equation}
f(\eta)=\left\{
\begin{array}
[c]{ll}%
\cos^{4}(\pi{}\eta), & |\eta{}|<1/2,\\[1mm]
0, & |\eta{}|>1/2.
\end{array}
\right.  \label{eq:fcos}%
\end{equation}
In a vicinity of each pulse, this function is close to the Gaussian
distribution
\begin{equation}
f_{\rm G}(\eta)=\exp\left(  -2\pi^{2}\eta^{2}\right)  \label{eq:fG}%
\end{equation}
in the interval, where $f_{\rm G}(\eta)$ is not small (see
Fig.~\ref{fig:pulse}).

In comparison with the Gaussian function (\ref{eq:fG}), function
(\ref{eq:fcos}) selected by us for the pulse simulation, on the one
hand, is more convenient for numerical calculations (the Gaussian
has to be artificially cut off beyond certain limits); on the other
hand, it corresponds to real pulse envelopes restricted in time. The
area of the pulse with the envelope described by function
(\ref{eq:fcos}) equals $\frac{3}{8}\Omega_{0}\tau$ or approximately
0.94 times the area of the corresponding Gaussian pulse. The
characteristic width of the latter equals $\tau_{\rm
G}\approx0.225\tau$.

\section{Hamiltonian}

\label{sec:Hamiltonian}Atom's state will be described by a wave function
constructed with the help of the Monte Carlo method \cite{Mol93}. After the averaging
over the ensemble of realized atomic states, this approach becomes equivalent
to the description of atom's state using the density matrix. At the same time,
in contrast to the latter, it allows an illustrative interpretation to be
given for the evolution of the state of a separate atom. The Hamiltonian, which
is used for the construction of such a wave function, differs from the
Hamiltonian used in the equation for the density matrix by a relaxation
term. It looks like
\begin{equation}
{H}=H_{0}+H_{\mathrm{int}}+H_{\mathrm{rel}}{,}\label{eq:Ham}%
\end{equation}
where the term
\begin{equation}
H_{0}=\hbar\omega_{0}|2\rangle\langle2|\label{eq:H0}%
\end{equation}
describes the atom in the absence of the field and the relaxation. The term
\begin{equation}
H_{\mathrm{int}}=-\bm{\mathrm{d}}_{12}|1\rangle\langle
2|\bm{\mathrm{E}}(t)-\bm{\mathrm{d}}_{21}|2\rangle\langle
1|\bm{\mathrm{E}}(t){,}\label{eq:Hint}%
\end{equation}
where $\mathbf{d}_{12}$ is the matrix element of the electric dipole moment for
the transition between states $|1\rangle$ and $|2\rangle$, which is
responsible for the atom-to-field interaction, and the term
\begin{equation}
H_{\mathrm{rel}}=-\frac{i\hbar\gamma}{2}|2\rangle\langle{}2|\label{eq:Hrel}%
\end{equation}
describes the relaxation owing to the spontaneous radiation emission. We
emphasize that the relaxation component of Hamiltonian
(\ref{eq:Ham}) looks like Eq.~(\ref{eq:Hrel}) if we
describe atom's state using the wave function constructed with
the help of the Monte Carlo method. However, if we use the density
matrix approach, the expression for $H_{\mathrm{rel}}$ is different
\cite{Mol93}.

\section{Atomic Wave Function}

\label{sec:WF}The solution of the Schr\"{o}dinger equation
\begin{equation}
i\hbar\frac{d}{dt}|\psi\rangle=H|\psi\rangle\label{eq:Sch}%
\end{equation}
is sought in the form
\begin{equation}
|\psi\rangle=c_{1}|1\rangle+c_{2}|2\rangle.\label{eq:psi}%
\end{equation}
The substitution of Eq.~(\ref{eq:Ham}) in Eq.~(z\ref{eq:Sch}) gives us the
following system of equations for $c_{1}$ and $c_{2}$:
\begin{equation}
\begin{array}{l}
 \displaystyle
i\hbar\frac{d}{dt}c_{1}=
-\bm{\mathrm{d}}_{12}\bm{\mathrm{E}}c_{2}{,}
 \\[5mm]
\displaystyle i\hbar\frac{d}{dt}c_{2}= \hbar\omega_{0}c_{2}-\bm{\mathrm{d}}_{21}%
\bm{\mathrm{E}}c_{1}-\frac{\gamma}{2}c_{2}.\!\!\!\!\!\!\!\! \\
\end{array}
\label{eq:c}%
\end{equation}
For further calculations, it is convenient to separate a rapidly
varying, with the frequency $\omega_{0}$, component in $c_{2}$. For
this purpose, we make the substitutions $c_{1}=C_{1}$ and
$c_{2}=C_{2}e^{-i\omega_{0}t}$. In the rotating wave approximation,
i.e. when we neglect the rapidly oscillating terms including
$e^{\pm2i\omega_{0}t}$ in the equations for $C_{1}$ and $C_{2}$,
Eq.~(\ref{eq:c}) yields
\begin{equation}
\begin{array}{l}
 \displaystyle\frac{d}{dt}C_{1}=-\frac{i}{2}(\Omega_{1}e^{-ikz+i\Phi_{1}}+\Omega_{2}
e^{ikz+i\Phi_{2}})C_{2}{,} \\[5mm]
\displaystyle\frac{d}{dt}C_{2}=-\frac{i}{2}(\Omega_{1}e^{ikz-i\Phi_{1}}
\!+\!\Omega_{2}e^{-ikz-i\Phi_{2}})C_{1}\!-\!\frac{\gamma}{2}C_{2}{,}\!\!\!\!\!\!\!\! \\
\end{array}
\label{eq:cc}
\end{equation}
where $\Omega_{1}=-\mathbf{d}_{12}\mathbf{e}E_{1}/\hbar$, $\Omega
_{2}=-\mathbf{d}_{21}\mathbf{e}E_{1}/\hbar$, $\delta=\omega_{0}-$
$-\omega$, $\Phi_{1}=\varphi_{1}-\delta t$, and
$\Phi_{2}=\varphi_{2}-\delta t$. Without loss of generality, the
Rabi frequencies $\Omega_{1}$ and $\Omega_{2}$ may be considered to
be real-valued quantities \cite{Sho90}.

Hamiltonian (\ref{eq:Ham}) is non-Hermitian, and the squared absolute value of
the wave function determined from the Schr\"{o}dinger equation changes in
time. In this connection, the procedure of function normalization should be
carried out after every small step in time. In addition, the condition of a
quantum jump within this time interval has to be testified \cite{Mol93}.

Let us take a wave function $|\psi(t)\rangle$ normalized to 1 at the time
moment $t$. The corresponding wave function $|\psi(t+\Delta{}t)\rangle$ at the
time moment $t+\Delta t$ can be found in two stages, as is described below
\cite{Mol93}.

\textbf{1}.~From the Schr\"{o}dinger equation (\ref{eq:Sch}), it follows
that, after a small enough $\Delta t$, the wave function
$|\psi(t)\rangle$ transforms into\vspace*{-2mm}
\begin{equation}
|\psi^{(1)}(t+\Delta{}t)\rangle\!=\!\left(\!1-\frac{i\Delta{}t}{\hbar}{{H}}\!\right)\!|\psi(t)\rangle.
\label{eq:phiI}
\end{equation}
Since Hamiltonian (\ref{eq:Ham}) is non-Hermitian,
$\psi^{(1)}(t+\Delta{}t)$ is not normalized to 1. The square of
its norm equals
\begin{equation}
\langle\psi^{(1)}(t+\Delta{}t)|\psi^{(1)}(t+\Delta{}t)\rangle=1-\Delta{}P{,}
\label{eq:phiIN}%
\end{equation}
where\vspace*{-2mm}
\begin{equation}
\Delta{}P=\frac{i\Delta{}t}{\hbar}\langle\psi(t)|H-H^{+}|\psi(t)\rangle
=\gamma\Delta{}t|C_{2}|^{2}. \label{eq:dPpsi}%
\end{equation}

\textbf{2}.~At the second stage, let us consider a possibility of
the quantum jump. If the value of random variable $\epsilon$, which
is uniformly distributed between zero and 1, is larger than $\Delta
P$~-- in most trials, it is the case, because $\Delta P\ll1$~-- the
jump is ignored, and the wave function at the time moment $t+\Delta
t$ is assumed \mbox{to equal}
\[
|\psi(t+\Delta{}t)\rangle=|\psi^{(1)}(t+\Delta{}t)\rangle/\sqrt{1-\Delta{}%
P},
\]\vspace*{-5mm}
\begin{equation}
\Delta{}P<\epsilon.
\end{equation}
However, if $\epsilon<\Delta P$, the jump takes place, and the atom goes into
the state
\begin{equation}
|\psi(t+\Delta{}t)\rangle=|1\rangle,\quad\Delta{}P>\epsilon.
\end{equation}

Now, let us apply this procedure to the case where the field does not act on
the atom (during the time interval between the light pulses). Let the initial
atom's state be
\begin{equation}
|\psi(0)\rangle=C_{1}(0)|1\rangle+C_{2}(0)|2\rangle.\label{eq:psi0}%
\end{equation}
At $t\rightarrow\infty$, the atom changes to a state $|1\rangle$ with the
probability equal to $|C_{1}(0)|^{2}$ with no photon emission or to
$|C_{2}(0)|^{2}$ if a photon is emitted.

If no quantum jump occurs within the time interval $[0,t]$, it follows from the
Schr\"{o}dinger equation with the \mbox{Hamiltonian}
\begin{equation}
{H}=H_{0}+H_{\rm rel}\label{eq:H0rel}%
\end{equation}
that\vspace*{-3mm}
\[
\psi^{(1)}(t)=\exp\left(\!-\frac{i}{\hbar}\int\limits_{0}^{t}H\,dt'\!\right)\psi(0)=
\]\vspace*{-5mm}
\begin{equation}
=C_{1}(0)|1\rangle+
C_{2}(0)\exp\left(\!-i\omega_{0}t-\frac{\gamma}{2}t\!\right)|2\rangle.
\label{eq:Sch-sol}
\end{equation}
Normalizing the function $\psi^{(1)}(t)$ to 1, we obtain
\begin{equation}
\psi(t)=C_{1}(t)|1\rangle+C_{2}(t)\exp\left(  -i\omega_{0}t)\right)
|2\rangle{,}\label{eq:Sch-sol-psi}%
\end{equation}
where
\begin{equation}
\begin{array}{l}
 \displaystyle
C_{1}(t)=\frac{C_{1}(0)}{\sqrt{|C_{1}(0)|^{2}+|C_{2}(0)|^{2}\exp(-\gamma
{}t)}}{,}
\\[5mm]
\displaystyle C_{2}(t)=
\frac{C_{2}(0)\exp\bigl(-\frac{1}{2}\gamma{}t\bigr)}{\sqrt
{|C_{1}(0)|^{2}+|C_{2}(0)|^{2}\exp(-\gamma{}t)}}.\!\!\!\!\!\!\!\! \\
\end{array}
\label{eq:Sch-ab}%
\end{equation}
The dependence of the wave function on the parameter $\gamma$, which is
inserted by the absence of spontaneous radiation emission within the time
interval $t$, is not evident. In the absence of radiation emission, the wave
function should seemingly be equal to
\begin{equation}
\psi(t)=C_{1}(0)|1\rangle+C_{2}(0)\exp(-i\omega_{0}t)|2\rangle.\label{eq:hyp}%
\end{equation}
In this case, the probability of radiation emission within the time
interval $[\Delta t{,}2\Delta t]$ would be the same as within the
time interval $[0{,}\Delta t]$. Moreover, this probability would be
the same for every of the following time intervals $\Delta t$,
because the population of the excited state, $|C_{2}(0)|^{2}$, remains
identical, and the atom would ultimately emit a photon. However,
this conclusion contradicts the circumstance that, if
$C_{1}(0)\neq0$, the atom is in the ground state with the
probability equal to $|C_{1}(0)|^{2}$ and will not emit a photon at
all \cite{Mol93}. At the same time, the indicated contradiction is
absent for the wave function (\ref{eq:Sch-sol-psi}), i.e. the
probability of photon emission falls down in the course of time and approaches
zero.

The probability that there will be no quantum jump within the time
interval $[0, t]$ equals \cite{Mol93}
\begin{equation}
P(t)=|C_{1}(0)|^{2}+|C_{2}(0)|^{2}\exp(-\gamma{}t){} \label{eq:P}%
\end{equation}
and agrees with the probability of the absence of a quantum jump, $|C_{1}(0)|^{2}$, at
$t\rightarrow\infty$ for the initial state (\ref{eq:psi0}) and with an exponential
decrease of the excited state population in the ensemble of atoms.

Therefore, atom's state at the time moment $t$ is described by the wave
function (\ref{eq:Sch-sol-psi}) with the probability $P(t)$ or by the wave
function
\begin{equation}
\psi(t)=|1\rangle\label{eq:psi-1-P}%
\end{equation}
with the probability $1-P(t)$.

\section{Atom's Motion}

\label{sec:Motion}Atom's motion will be described in the framework of
classical mechanics. During its interaction with light pulses, the atom
undergoes the action of the force \cite{Min86}
\begin{equation}
F=(\varrho_{12}\bm{\mathrm{d}}_{21}+\varrho_{21}\bm{\mathrm{d}}_{12})\frac{\partial\bm{\mathrm{E}}}{\partial z}{,} \label{eq:F}%
\end{equation}
where the density matrix elements are expressed in terms of $C_{1}$ and
$C_{2}$ as follows:
\begin{equation}
\begin{array}{l}
 \displaystyle
\varrho_{11}=|c_{1}|^2=|C_{1}|^2{,}
\\[2mm]
\varrho_{22}=|c_{2}|^2=|C_{2}|^2{,}
\\[2mm]
\varrho_{12}=c_{1}c_{2}^{*}=C_{1}C_{2}^{*}e^{i\omega_{0}{}t}{,}
\\[2mm]
\varrho_{21}=c_{2}c_{1}^{*}=C_{2}C_{1}^{*}e^{-i\omega_{0}{}t}.
\end{array}
\label{eq:rho}
\end{equation}

After the averaging over the period of oscillations with the frequency
$\omega _{0}$, the expression for force (\ref{eq:F}) in field
(\ref{eq:E}) reads
\begin{equation}
{F}=\hbar{}k\mathop{\mathrm{Im}}C_{1}C_{2}^{\ast}(  \Omega_{1}%
e^{ikz-i\Phi_{1}}-\Omega_{2}e^{-ikz-i\Phi_{2}}). \label{eq:FF}%
\end{equation}
The dependences of atom's coordinate $z$ and velocity $v$ on the time are
determined from the equations
\begin{equation}
\dot{v}=F/m{,}\label{eq:ma}
\end{equation}\vspace*{-7mm}
\begin{equation}
\dot{z}=v{,} \label{eq:dz}
\end{equation}
where $m$ is atom's mass.

Proceeding from the probability of a quantum jump equal to $1-P(t)$,
where $P(t)$ between light pulses is determined by formula
(\ref{eq:P}), we simulate the time moment of spontaneous radiation
emission for every realization of the wave function. At this moment,
atom's coordinate does not change, and the projection of the
momentum $\hbar\mathbf{k}$ transferred to the atom along the
$z$-axis is determined, by assuming a random direction of light
quantum \mbox{emission.}

\section{Numerical Calculation Routine}

\label{sec:num}To simulate the motion of atom, Eqs.~(10),
(\ref{eq:ma}), and (\ref{eq:dz}) were solved simultaneously. The
time intervals, in which the atom interacts with the field, were
divided into small subintervals, where the wave function was
normalized and the presence of a quantum jump was checked. If so,
atom's velocity was modified according to the formula
\begin{equation}
\Delta{}v=\hbar{}k\cos(\pi\epsilon){,}\label{eq:vsp}%
\end{equation}
where $0<\epsilon<1$ is a random number with a uniform distribution over the
interval $[0,1]$.

For the time intervals, when the field does not act on the atom, the wave function
can be written down in an analytical form. Therefore, the calculation time can
be considerably reduced, if, instead of simulating the quantum jump within
numerous short time intervals during the free evolution of the atom, we will
determine at once if there was a jump within the whole interval of free atom's
evolution and, if so, at which moment it happened. Let the atom after its
interaction with the field be described by the wave function (\ref{eq:psi0}). The
value of random variable $\epsilon$ should be compared with $|C_{1}(0)|^{2}$.
The jump takes place if $\epsilon>|C_{1}(0)|^{2}$, and does not otherwise. Let
us simulate the time moment of a quantum jump. We take another value for
$\epsilon$ and calculate $\ln\epsilon$. For the exponential distribution of
the probability
\begin{equation}
P_{e}=e^{-\gamma{}t}{,}\label{eq:Pe}%
\end{equation}
the quantity
\begin{equation}
t_{\mathrm{jump}}=-\left(  \ln\epsilon\right)  /\gamma\label{eq:tj}%
\end{equation}
simulates the time moment, when the jump happens~\cite{Sob73}.

\begin{figure}
\vskip1mm
\includegraphics[width=7.2cm]{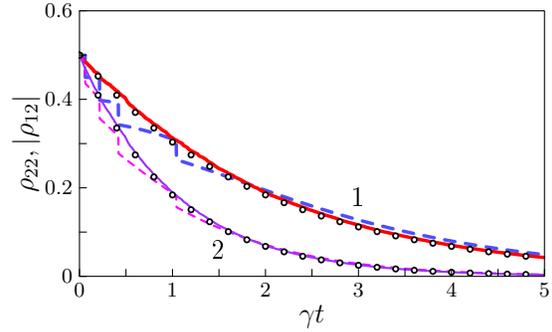}
\vskip-3mm\caption{Comparison of the density matrix elements
$\rho_{12}$ (curves~\textit{1}) and $\rho_{22}$ (curves~\textit{2})
calculated from the Monte Carlo wave functions (the dashed curves
correspond to the averaging over 10 and the solid curves over
1000~realizations) with the exact expression (\ref{eq:rho-sol}) for
the density matrix (circles). The initial conditions are
$\rho_{12}(0)=1/2$, $\rho_{22}(0)=1/2$, and $|\psi(0)\rangle=\sqrt
{2}\left[ |1\rangle+|2\rangle\right]  /2$  }\label{fig:MCWF}
\end{figure}

\section{Example of Density Matrix Calculation Using the Wave Function
Constructed with~the Help of the Monte Carlo Method}

\label{sec:free}Let us illustrate the equivalence between the descriptions of
an atomic ensemble with the use of the wave function constructed with the help
of the Monte Carlo method and the density matrix. Consider the free evolution of
an atom, for which the variation of the density matrix in time can be easily
calculated. The equation for the density matrix evolution looks like
\[
\frac{d}{dt}\rho=i\omega_{0}[\rho,|2\rangle\langle{}2|]-
\]\vspace*{-7mm}
\begin{equation}
-\frac{\gamma}{2}\left(|2\rangle\langle{}2|\rho+\rho{}|2\rangle\langle{}2|\right)
+\gamma{}|1\rangle\langle{}2|\rho{}|2\rangle\langle{}1|).
\label{eq:Hrho}
\end{equation}
From whence, the equations for the density matrix elements follow,
\begin{equation}
\begin{array}{l}
 \displaystyle
\frac{d}{dt}\rho_{22}=-\gamma\rho_{22},
\\[5mm]
\displaystyle
\frac{d}{dt}\rho_{12}=i\omega_{0}-\frac{\gamma}2{}\rho_{12},
\\[5mm]
\displaystyle \rho_{21}=\rho_{12}^{*},\quad{}\rho_{11}=1-\rho_{22},
\end{array}
\label{eq:drho}
\end{equation}
and we obtain
\begin{equation}
\begin{array}{l}
 \displaystyle
\rho_{22}(t)=\rho_{22}(0)\exp(  -\gamma{}t)  ,
\\[5mm]
\displaystyle \rho_{12}(t)=\rho_{12}(0)\exp\left(\!
i\omega_{0}t-\frac{1}{2}\gamma {}t\!\right)\!.
\end{array}
\label{eq:rho-sol}%
\end{equation}
The density matrix elements can also be calculated using the wave
function constructed with the help of the Monte Carlo method,
Eq.~(\ref{eq:rho}), and, afterward, carrying out the averaging over the
ensemble of wave function realizations. Figure~\ref{fig:MCWF}
testifies that the agreement between the density matrix elements
calculated with the use of both approaches is satisfactory in the case
of 10~wave function realizations and excellent in the case of
1000~ones.

\begin{figure}
\vskip1mm
\includegraphics[width=7.5cm]{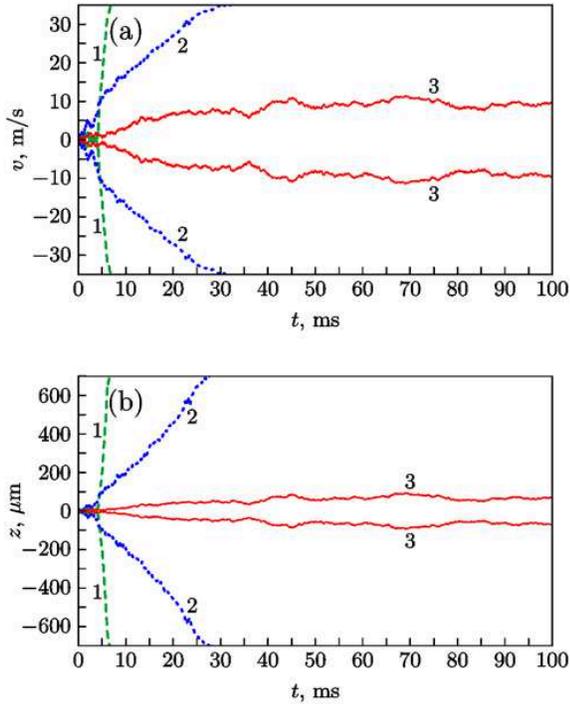}
\vskip-3mm\caption{Time evolution of intervals (between the upper
and lower curves with identical numbers), in which the velocity
(\textit{a}) and the coordinate (\textit{b}) of an atom with the
mass $m=40~\mathrm{a.m.u.}$ in the trap created by the fields of
counter-propagating $\pi$-pulses change. The pulse duration
$\tau=1$~\textrm{ps}, the pulse repetition period
$T=10$~\textrm{ns}, the carrier frequency of light pulses
corresponds to a wavelength of 600\textrm{~nm}, and
$\varphi_{1}=\varphi_{2}=0$. The parameter $\gamma=2\pi\times10^{6}$
(\textit{1}), $2\pi\times10^{7}$ (\textit{2}), and
$2\pi\times10^{8}~\mathrm{s}^{-1}$ (\textit{3}). The initial
conditions are $z=0$, and $v=0.2~\mathrm{m/s}$  }\label{fig:pi-40}
\end{figure}

\section{Motion of Atoms and Nanoparticles in~a~Trap Formed by Sequences of~Counter-Propagating \boldmath$\pi$-Pulses}

\label{sec:pi}Consider atom's motion in a trap formed by sequences
of counter-propagating $\pi$-pulses. We suppose that the field
frequency is resonant with that of a transition in the atom,
$\delta=0$. When explaining the mechanism of trap action, we
proceeded from the assumption that the atom at point $A$ (see
Fig.~\ref{fig-scheme}) is in the ground state. Then, the sequence of
influences by pulses $R$ and $L$ results in a variation of its
momentum by $2\hbar k$ toward the trap center, point $C$. It is true
in most cases. Really, since the time interval between pulses $L$
and $R$ is much shorter than that between pulses $L$ and
$R^{\prime}$, the probability of spontaneous radiation emission by
the atom in the excited state is much higher in the latter case. As
a result, the atom at point $A$ will predominantly be in the ground
state, and the force of light pressure on it will be directed toward
the trap center. However, if the atom goes into the ground state
after the action of pulse~$R$ owing to spontaneous radiation
emission, pulse~$L$ returns it back to the excited state, and,
during some time before the spontaneous emission, the atom undergoes
the action of the force directed from the trap center. This process,
when the force changes its direction, gives rise to a momentum
diffusion, i.e. a spread of the atomic distribution with respect to
atomic \mbox{momenta \cite{Voi91,Neg08}.}\looseness=1

\begin{figure}
\vskip1mm
\includegraphics[width=7.5cm]{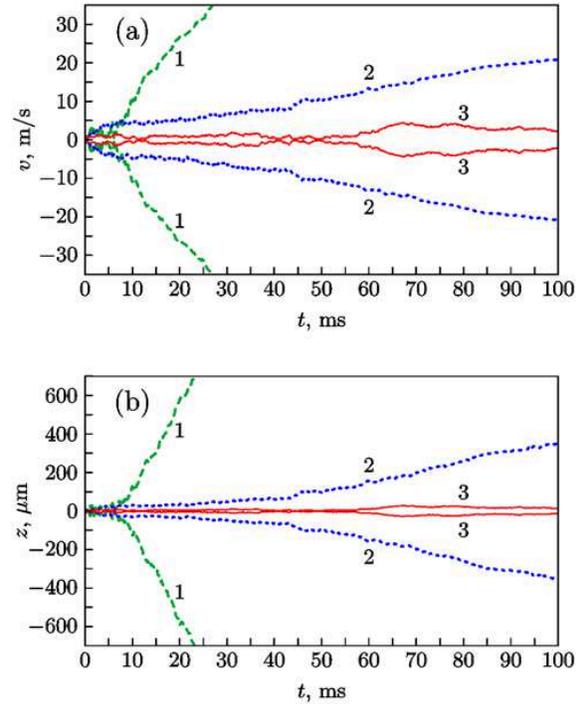}
\vskip-3mm\caption{The same as in Fig.~\ref{fig:pi-40}, but for an
atom with the mass $m=200~\mathrm{a.m.u.}$  }\label{fig:pi-200}
\end{figure}

It is evident that the momentum diffusion process substantially
depends on the rate of spontaneous radiation emission $\gamma$. In
particular, the smaller is the value of $\gamma T$, the larger
number of pairs of light pulses change atom's momentum in the
direction from point $C$ before the spontaneous radiation event. The
momentum diffusion is minimal at $\gamma T\gg1$, when the atom in
the excited state at point $A$ has enough time to go into the ground
state before the arrival of pulse $R^{\prime}$.
Figures~\ref{fig:pi-40} and~\ref{fig:pi-200} illustrate a reduction
of the momentum diffusion effect with the growth of $\gamma T$. They
exhibit the intervals of changes in the velocity and the coordinate
with time for atoms with the masses $m=40$\ and 200~a.m.u.,
respectively, and for various rates of spontaneous radiation
emission by the atom in the excited state. The time dependences of
the coordinate and the velocity themselves are not shown, because
the atom oscillates between the upper and lower curves marked by the
same number, and the oscillation period is so small (less than
0.2~ms for the dependences shown in the figures) that the curves
describing those dependences would entirely fill the space between
the limits indicated in \mbox{the figures.}\looseness=1

At $\gamma T<1$, the time dependences of the interval, in which atom's
velocity varies (curves~\textit{1} and \textit{2} in Figs.~\ref{fig:pi-40}
and~\ref{fig:pi-200}), only vaguely resemble the root one, which is typical of
diffusion processes. The earlier researches \cite{Voi91,Neg08} of the momentum
diffusion in the field of counter-propagating $\pi$-pulses and for a fixed
delay between the pulses showed that the momentum diffusion coefficient
\begin{equation}
D=\frac{1}{2}\lim\limits_{t\rightarrow\infty}\frac{\langle{}p^{2}%
\rangle-\langle{}p\rangle^{2}}{t}, \label{eq:D}%
\end{equation}
where $p$ is atom's momentum, and the notation $\left\langle
\ldots\right\rangle $ means the averaging over the ensemble, is approximately
proportional to the delay between counter-propagating pulses. In our case,
when the atom moves near the point, where those pulses \textquotedblleft
collide\textquotedblright, this delay is proportional to the atomic coordinate
that changes in time. Hence, a simple analytical expression for $D$ obtained
in work \cite{Voi91} is unsuitable in our case, and we can judge the
character of diffusion-induced variations in the atomic momentum only on the
basis of numerical calculations.

When the quantity $\gamma T$ increases above 1 (curves~\textit{3}),
the atom, owing to a high probability of its relaxation between
pairs of counter-propagating pulses that are close in time, is
practically always in the ground state before its interaction with
the field.\,\,This case was analyzed in work \cite{Fre95} for a
one-time interaction between the atom and the field of a pulse pair,
when the width of an atomic wave packet is wide in comparison with
the wavelength of laser radiation.\,\,In the region where
counter-propagating pulses overlap, the field is close to that of a
standing wave, and the diffraction of a wave packet takes place,
which results in a spreading of the atomic distribution with respect
to atomic momenta.\,\,If the atomic wave packet is narrow in
comparison with the wavelength of laser radiation, which corresponds
to the classical description of atom's motion, the distribution
spreading over atomic momenta occurs owing to the pulsed
atom-to-field interaction in the region, where the light pulses
spatially overlap--the momentum transferred to the atom is
determined by its relative coordinate with respect to the maxima and
the minima of the field formed by counter-propagating laser
pulses.\,\,The atom does not move periodically in the
field.\,\,Therefore, when crossing the region where the light pulses
overlap, it interacts every time with the field at a different point
and, accordingly, obtains different momenta from the field. As a
result, the character of its momentum change approaches the chaotic
one (see curves~\textit{3} in Figs.~\ref{fig:pi-40}
and~\ref{fig:pi-200}).\,\,Certainly, this mechanism of interaction
between the atom and the field is valid in the case $\gamma T\ll1$
as well, but the momentum variation owing to a change of the force
direction after the event of spontaneous radiation emission
\mbox{dominates here}.\looseness=2

When comparing Figs.~\ref{fig:pi-40} and~\ref{fig:pi-200}, one can
see that, under identical initial conditions, the oscillation limits
of both the velocity and the coordinate get rapidly narrower for
heavier atoms.\,\,This fact is associated with a reduction of atom's
velocity variation $v_{r}=\hbar k/m=h/(\lambda m)$, where $\lambda$\
is the wavelength of laser radiation, when absorbing or emitting a
photon ($v_{r}=1.7~\mathrm{cm/s}$ for $m=40~\mathrm{a.m.u.}$ and
3.3~mm/s for $m=200~\mathrm{a.m.u.}$), which is analogous to a
reduction of the diffusion coefficient in a gas for shorter mean
\mbox{free paths.}

Hence, the trap created on the basis of counter-propagating light
pulses possesses better properties for holding the heavy atoms~--
probably, with an additional cooling field \cite{Fre95}~-- such as
Rb, Cs, and Th. Pulses with a short duration in comparison with the
period of their repetition weakly perturb the atomic state. This is
important, e.g., for high-precision spectroscopic researches, in
particular, for the implementation of the frequency standard based
on the optical nuclear transition in thorium \cite{Pei03,Pei09}. The
trap has even better prospects for holding the nanoparticles with a
low content~-- e.g., 0.1\%~-- of \textquotedblleft
active\textquotedblright\ atoms with the transition frequency close
to the carrier frequency of laser pulses. Such nanoparticles behave
as \textquotedblleft heavy atoms\textquotedblright\ with the atomic
masses equal to tens of thousands. Accordingly, the momentum
diffusion for such nanoparticles should be slower; therefore, they
can be held in the trap using no additional \mbox{cooling field.}

\begin{figure}
\vskip1mm
\includegraphics[width=7.6cm]{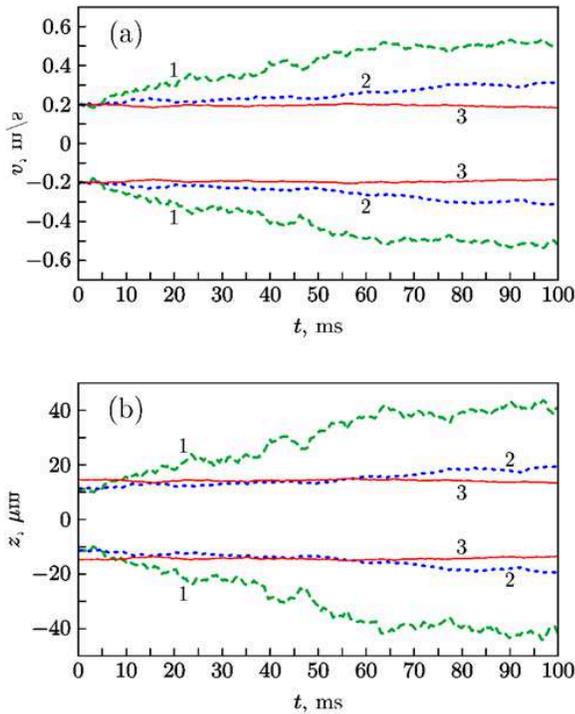}
\vskip-3mm\caption{The same as in Fig.~\ref{fig:pi-40}, but for a
nanoparticle with the specific mass $m=30000~\mathrm{a.m.u.}$ per
one \textquotedblleft active\textquotedblright\ atom. The $z$-axis
is directed vertically  }\label{fig:pi-nano}
\end{figure}

Figure~\ref{fig:pi-nano} demonstrates the variation intervals for the velocity
and the coordinate of a nanoparticle with the specific mass
$m=30000~\mathrm{a.m.u.}$ per one \textquotedblleft active\textquotedblright%
\ atom. The curves were plotted for various spontaneous radiation
rates of an atom in the excited state. The $z$-axis is directed
upward. One can see that the gravitation force practically does not
affect the motion of a nanoparticle. Similarly to the atoms with the
masses $m=40$ and 200$~\mathrm{a.m.u.}$, the momentum diffusion
decreases with the growth of $\gamma T$. In the case of large
$\gamma T$ (curves~\textit{3}), the oscillation limits for the
nanoparticles do not change in time, at least till 0.1~s, and the
influence of the momentum diffusion on nanoparticle's motion is weak
even at $\gamma T\ll1$ (curves~\textit{1} and \textit{2}). Since the
rate of spontaneous radiation emission $\gamma$ is fixed in every
specific case, the product $\gamma T$ can be changed as required by
varying the pulse repetition sequence. In the calculations carried
out for the parameters that correspond to Figs.~\ref{fig:pi-40} and
\ref{fig:pi-200}, but at $T=10^{-7}~\mathrm{s}$, a considerable
reduction of the diffusion influence on atoms' motion was
\mbox{observed.}\looseness=1

Another controllable parameter, on which the interaction between the atom and the
field depends, is a detuning of the light pulse carrier frequency from the
frequency of the transition in the atom. In the next section, we shall demonstrate
that even an insignificant, in comparison with the Rabi frequency, detuning of
light pulses substantially modifies atom's motion in the trap; in
particular, it can suppress the momentum diffusion, provided a proper choice
of other parameters.

\begin{figure}
\vskip1mm
\includegraphics[width=7.5cm]{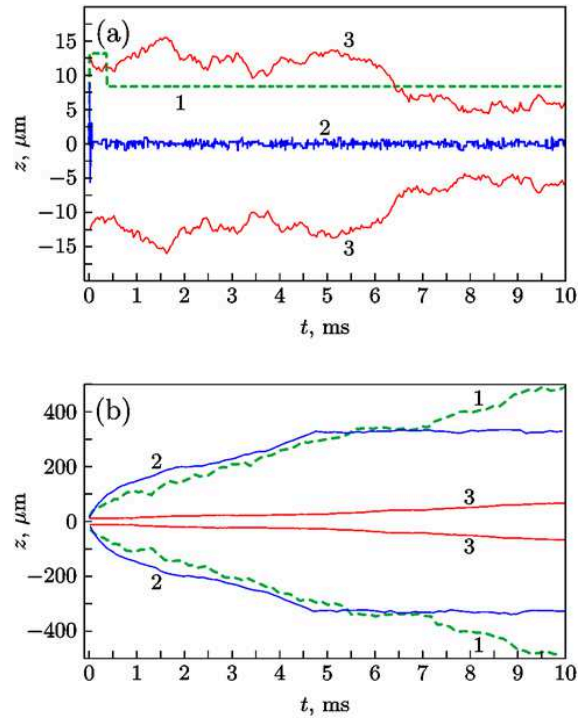}
\vskip-3mm\caption{Time dependences of the coordinate and intervals
of its changes (between the upper and lower curves with identical
numbers) for an atom with the mass $m=200~\mathrm{a.m.u.}$ in the
field of counter-propagating $\pi$-pulses. The pulse duration
$\tau=1$~\textrm{ps}, the pulse repetition period
$T=10$~\textrm{ns}, and $\varphi_{1}=\varphi_{2}=0$. The carrier
frequency of light pulses corresponds to a wavelength of
600\textrm{~nm} and is detuned by $10^{10}~\mathrm{s}^{-1}$ below
(\textit{a}) and above (\textit{b}) the transition frequency. The
parameter $\gamma=2\pi \times10^{6}$ (\textit{1}),
$2\pi\times10^{7}$ (\textit{2}), and $2\pi
\times10^{8}~\mathrm{s}^{-1}$ (\textit{3}). The initial conditions
are $z=0$,
and $v=2~\mathrm{m/s}$%
  }\label{fig:det}\vspace*{-2mm}
\end{figure}

\section{Atom's Motion in a Trap Formed by~Sequences of Counter-Propagating
Pulses That Are Non-Resonant with the Transition Frequency in the
Atom}

\label{sec:det}Consider how the detuning of the laser radiation frequency from
the resonance with the atomic transition frequency $\omega_{0}$ affects the
interaction between the atom and the sequence of counter-propagating laser
pulses. Figure~\ref{fig:det} illustrates an example of the time dependence of
the atomic coordinate (curves~\textit{1}\ and \textit{2} in panel~\textit{a}) and
shows the upper and lower coordinate limits, between which
the atom oscillates (all other curves). One can see that, if $\delta>0$ (i.e.
the carrier frequency of pulses is lower than the transition frequency in
the atom) and $\gamma T<1$, the atom, after a short transient process, becomes
localized in a narrow spatial interval. At $\gamma T>1$, the atom oscillates
with a varying amplitude, by demonstrating a tendency to a reduction. This means
that, if the atom is localized, the preservation of the coherent character of
an atomic state within the pulse repetition period plays an essential role. On
the other hand, at a low spontaneous radiation rate $\gamma=2\pi\times10^{7}%
$~s$^{-1}$ (not shown in the figure), the amplitude of oscillations
grows. This fact agrees well with the expression obtained in work
\cite{Voi91} for the momentum diffusion coefficient in the case
$\gamma T\ll1$, namely, $D\sim1/(\gamma T)$. Therefore, as the
quantity $\gamma T$ decreases, the growth of the oscillation
amplitude owing to the momentum diffusion should expectedly dominate
over the atomic deceleration in the field of
counter-propa\-gating~pulses.\looseness=1

\begin{figure}
\vskip1mm
\includegraphics[width=\column]{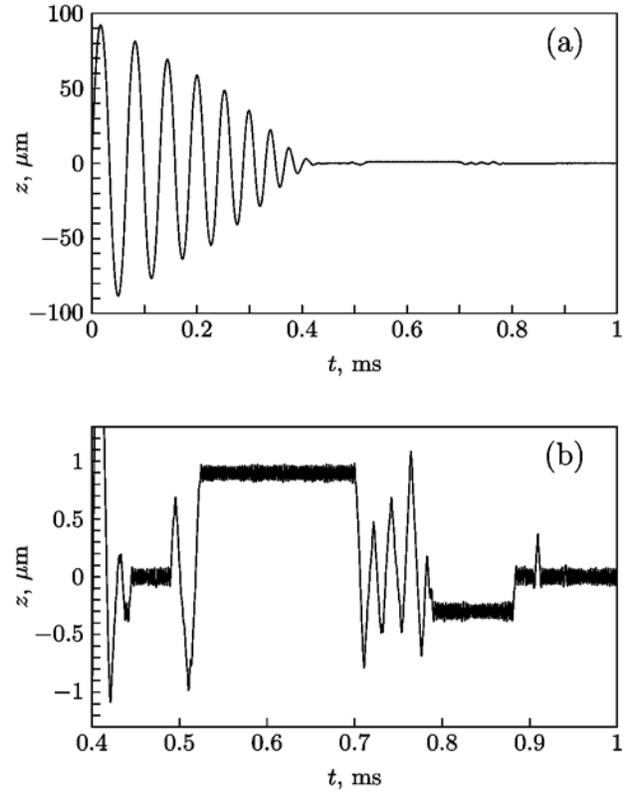}
\vskip-3mm\caption{Time dependence of the coordinate for an atom
with the mass $m=200~\mathrm{a.m.u.}$ in the field of
counter-propagating $\pi$-pulses (\textit{a}) and its scaled-up
section (\textit{b}). The pulse duration $\tau=1$~\textrm{ps}, the
pulse repetition period $T=10$~\textrm{ns}, and $\varphi_{1}=$
$=\varphi_{2}=0$. The carrier frequency of light pulses corresponds
to a wavelength of 600\textrm{~nm} and is by
$10^{10}~\mathrm{s}^{-1}$ below the transition frequency.\ The rate
of excited state damping $\gamma =2\pi\times10^{7}~\mathrm{s}^{-1}$.
The initial conditions are $z=0$, and $v=10~\mathrm{m/s}$
}\label{fig:det-i}
\end{figure}

If the detuning sign changes (Fig.~\ref{fig:det}), the amplitude of atomic
oscillations either grows (curves~\textit{1} and~\textit{3}) or saturates
(curve~\textit{2}).

We made calculations for a number of other realizations of the Monte
Carlo wave function, with other phases $\varphi_{1}$ and
$\varphi_{2}$, and with a pulse area of $0.8\pi$ and obtained the
time dependences of atom's coordinate that are analogous to those
exhibited in Fig.~\ref{fig:det}. The dependences of the same type
were obtained for $\left\vert \delta\right\vert =10^{9}$ and
$5\times10^{10}$~s$^{-1}$. At the same time, for $\left\vert
\delta\right\vert =10^{7}$, 10$^{8}$, and $10^{11}$~s$^{-1}$, the
decrease of atom's velocity and its localization is observed at the
opposite detuning, i.e. at $\delta<0$. The last result agrees with
that obtained in work \cite{Mol91}, where, as $\delta$ varied
($\delta>0$), the cooling of atoms is changed by their heating. If
the detuning diminished to $\left\vert \delta\right\vert
=10^{6}$~s$^{-1}$, no velocity reduction was observed for the
parameters in Fig.~\ref{fig:det}. The results of numerical
simulation of atom's motion in the field of sequences of
counter-propagating pulses brings us to a conclusion that a
reduction of the amplitude and, accordingly, the velocity of atomic
oscillations in the trap is possible in a wide interval of the field
carrier frequency detuning from the resonance with the frequency of
the atomic transition, provided the proper selection of
$\delta$-sign. The pulse repetition period should be small enough
for the condition $\gamma T<1$ to be \mbox{satisfied.}\looseness=1

Figure~\ref{fig:det-i} illustrates the time dependence of the
coordinate for parameters that provide the deceleration of an atom
in the trap. At calculations, the initial velocity of the atom was
selected to be much higher than that in Fig.~\ref{fig:det}. Owing to
the interaction of the atom with light pulses, its velocity
decreases from 10 to 0.2~m/s, which corresponds to the temperature
change for an ensemble of 200-a.m.u. atoms with an initial
root-mean-square velocity of 10~m/s from 2.4~K to 1~mK during
0.5~ms. The limiting temperature is a little higher that the Doppler
cooling limit for atoms, $\hbar\gamma/k_{\rm
B}\approx0.4~\mathrm{mK}$, where $k_{\rm B}$ is the Boltzmann
constant. Since the variation of the atomic velocity at the
absorption or emission of a photon, $v_{\rm rec}=\hbar k/m$, equals
approximately 0.003~m/s, at least 3000~photons were absorbed and
emitted. As is seen from Fig.~\ref{fig:det-i},\textit{b}, the atom
oscillates near the antinodes of a light wave, at a short distance
from the coordinate origin, much shorter than the light pulse
extension in the space (in particular, around $z=0$, $-\lambda/2$,
$3\lambda/2$). At such distances from the coordinate origin, the
fields of pulses from counter-propagating waves are almost
identical. As a result, the atom is subjected to the action of a
sequence of standing wave pulses, and cold atoms become captured at
the corresponding antinodes. It is a well-known phenomenon that is
observed in the stationary field of a monochromatic standing
\mbox{wave \cite{Pav93}.}\looseness=1

\section{Conclusions}

\label{sec:conclu}In this work, we have demonstrated that the
momentum diffusion (a spread of the momentum distribution in an
ensemble of atoms) in the resonance field of sequences of
counter-propagating $\pi$-pulses or pulses close to them, which
create a trap for atoms, depends substantially on the ratio between
of the pulse repetition period and the time of spontaneous radiation
emission.\,\,At $\gamma T\sim1$, the atom can be held in the trap
for not less than 0.1~s.\,\,While calculating the motion of
nanoparticles in the trap under the condition $\gamma T\sim1$, we
did not observe the growth of the amplitude of their oscillatory
motion around the \mbox{trap center.}\looseness=1

For a non-resonance field, we can made a selection of the detuning
of light pulse carrier frequency and $\gamma T$ such that the
amplitude of atom's oscillations in the trap would considerably
decrease, and the temperature in an ensemble of atoms could decrease
almost to the Doppler cooling limit. As a result, the atoms become
localized in vicinities of the antinodes of the non-stationary
standing wave that is formed by counter-propagating light pulses in
the region around the point of their \textquotedblleft
collision\textquotedblright. Since the field acts on the atom only
during a short time interval, the pulse-created light trap can be
used for high-precision spectroscopic researches. In particular, it
can be used to hold atoms or ions of thorium-229, while developing
the frequency standard on the basis of an optical nuclear transition
\cite{Pei03,Pei09}. Another possible application of the trap on the
basis of counter-propagating light pulses can be the manipulation
with nanoparticles containing a small fraction of atoms (of about
0.1\% or smaller) with the transition frequency close to the carrier
frequency of \mbox{light pulses.}\looseness=1

\vskip3mm

{\it The work was executed in the framework of the State
goal-oriented scientific and engineering program \textquotedblleft
Nanotechnologies and Nanomaterials\textquotedblright\ (themes
N~1.1.4.13 and 3.5.1.24).and sponsored by the State Fund for
Fundamental Researches of Ukraine (project No.~F40.2/039).}

\rezume{%
В.І.~Романенко, О.В.~Романенко,\\ О.Г.~Удовицька,
Л.П.~Яценко}{ІМПУЛЬСНА ДИФУЗІЯ\\ АТОМІВ І НАНОЧАСТИНОК У ОПТИЧНІЙ\\
ПАСТЦІ, УТВОРЕНІЙ ПОСЛІДОВНОСТЯМИ\\ ЗУСТРІЧНИХ СВІТЛОВИХ ІМПУЛЬСІВ}
{Розглянуто рух атомів і наночастинок у пастці, утвореній
послідовностями  зустрічних світлових імпульсів. Стан атома
описується хвильовою функцією, побудованою за допомогою методу
Монте-Карло, його рух~-- класичною механікою. Оцінено вплив
імпульсної дифузії, зумовленої спонтанним випромінюванням збуджених
атомів та імпульсним характером взаємодії атома з полем, на рух
атома чи наночастинки у пастці. Показано, що при належно обраних
параметрах взаємодії атома з полем рух атома у пастці гальмується, і
він коливається біля пучностей нестаціонарної стоячої хвилі,
сформованої зустрічними світловими імпульсами поблизу точки, де вони
``зіштовхуються''.}


\begin{thebibliography}{99}                                                                                               %
\bibitem {Min86}V.G.~Minogin and V.S.~Letokhov, \textit{Laser Light Pressure
on Atoms} (Gordon and Breach, New York, 1987).\vspace*{0.5mm}
\bibitem {Kaz91}A.P.~Kazantsev, G.I.~Surdutovich, and V.P.~Yakovlev,
\textit{Mechanical Action of Light on Atoms} (World Scientific,
Singapore, 1990).\vspace*{0.5mm}
\bibitem {Pav93}B.D.~Pavlik, \textit{Cold and Ultracold Atoms} (Naukova Dumka,
Kiev, 1993) (in Russian).\vspace*{0.5mm}
\bibitem {Met99}H.J. Metcalf and P. van der Stratten, \textit{Laser Cooling and
Trapping} (Springer, Berlin, 1999).\vspace*{0.5mm}
\bibitem {Chu98}S. Chu, Rev. Mod. Phys. \textbf{70}, 685 (1998).\vspace*{0.5mm}
\bibitem {Coh98}C.N. Cohen-Tannoudji, Rev. Mod. Phys. \textbf{70}, 707
(1998).\vspace*{0.5mm}
\bibitem {Phi98}W.D. Phillips, Rev. Mod. Phys. \textbf{70}, 721
(1998).\vspace*{0.5mm}
\bibitem {Bal00}V.I. Balykin, V.G. Minogin, and V.S.~Letokhov, Rep. Prog.
Phys. \textbf{63}, 1429 (2000).\vspace*{0.5mm}
\bibitem {Fre95}T.G.M. Freegarde, J. Waltz, and W.~H\"{a}nsch, Opt. Commun.
\textbf{117}, 262 (1995).\vspace*{0.5mm}
\bibitem {Goe97}A. Goepfert, I. Bloch, D.~Haubrich, F.~Lison, R.~Sch\"{u}\-tze,
R.~Wynands, and D.~Meshede, Phys. Rev.~A \textbf{56}, R3345
(1997).\vspace*{0.5mm}
\bibitem {Bal05}V.I.~Balykin, JETP Lett. \textbf{81}, 206 (2005).\vspace*{0.5mm}
\bibitem {Rom11}V.I.~Romanenko and L.P.~Yatsenko, J. Phys.~B \textbf{44},
115305 (2011).\vspace*{0.5mm}
\bibitem {Rom12}V.I.~Romanenko and L.P.~Yatsenko, Ukr. Fiz. Zh. {\bf 57}, 893 (2012).\vspace*{0.5mm}
\bibitem {Kaz74}A.P. Kazantsev, Sov. Phys. JETP \textbf{39} 784 (1974).\vspace*{0.5mm}
\bibitem {Voi91}V.S.~Voitsekhovich, M.V.~Danileiko, A.M.~Negriyko,
V.I.~Ro\-manenko, L.P.~Yatsenko, Zh. \`{E}ksp. Teor. Fiz.
\textbf{99}, 393 (1991).\vspace*{0.5mm}
\bibitem {Neg08}A.M.~Negriyko, V.I.~Romanenko, and L.P.~Yatsenko,
\textit{Dynamics of Atoms and Molecules in Coherent Laser Fields}
(Naukova Dumka, Kyiv, 2008) (in Ukrainian).\vspace*{0.5mm}
\bibitem {Let76}V.S. Letokhov, V.G. Minogin, B.D.~Pavlik, Opt. Commun.
\textbf{19}, 72 (1976).\vspace*{0.5mm}
\bibitem {Let77}V.S. Letokhov, V.G. Minogin, and B.D. Pavlik, Sov. Phys. JETP
\textbf{45}, 698 (1977).\vspace*{0.5mm}
\bibitem {Mol91}K. M{\o }lmer, Phys. Rev. Lett. \textbf{66}, 2301 (1991).\vspace*{0.5mm}
\bibitem {Mol93}C. M{\o }lmer, Y. Castin, and J.~Dalibard, J.~Opt. Soc. Am.~B
\textbf{10}, 524 (1993).\vspace*{0.5mm}
\bibitem {Ber98}K. Bergmann, H. Theur, and B.W.~Shore, Rev. Mod. Phys.
\textbf{70}, 1003 (1998).\vspace*{0.5mm}
\bibitem {Vit01}N.V. Vitanov, T. Halfmann, B.W.~Shore, and K.~Berg\-mann, Annu.
Rev. Phys. Chem. \textbf{52}, 763 (2001).\vspace*{0.5mm}
\bibitem {Rom06}V.I. Romanenko, Ukr. J. Phys. \textbf{51}, 1054 (2006).\vspace*{0.5mm}
\bibitem {Sho90}%
B.W.~Shore, \textit{The Theory of Coherent Atomic Excitation,
Vol.~1: Simple Atoms and Fields} (Wiley, New York,
1990).\vspace*{0.5mm}
\bibitem {Sob73}I.M.~Sobol, \textit{The Monte Carlo Method} (Univ. of
Chicago Press, Chicago, IL, 1974).\vspace*{0.5mm}
\bibitem {Pei03}E. Peik and Ch. Tamm, Europhys. Lett. \textbf{61}, 181 (2003).\vspace*{0.5mm}
\bibitem {Pei09}E. Peik, K. Zimmermann, M.~Okhapkin, and Ch. Tamm, in
\textit{Proceedings of the 7th Symposium on Frequency Standards and
Metrology,} edited by L.~Maleki (World Scientific, Singapore, 2009),
p.~532. \vspace*{2mm}

\begin{flushright}
{\footnotesize Received 27.08.12.\\ Translated from Ukrainian by
O.I.~Voitenko}
\end{flushright}
\end{thebibliography}
\end{document}